\documentclass[11pt]{article}
\usepackage{amsthm,amsmath,epsf,amssymb,enumerate,array,color}
\usepackage[all,matrix,arrow,import,arc,poly,color,ps,dvips]{xy}

\definecolor{dblue}{rgb}{0,0,.7} \definecolor{indigo}{RGB}{50,0,105}

\newtheorem{thm}{Theorem}[section]        \newtheorem{lemma}[thm]{Lemma}	\newtheorem{cor}[thm]{Corollary}
\newtheorem{definition}[thm]{Definition} \newtheorem{prop}[thm]{Proposition}

\DeclareFontFamily{U}{rsf}{} \DeclareFontShape{U}{rsf}{m}{n}{  <5> <6> rsfs5 <7> <8> <9> rsfs7 <10-> rsfs10}{}
\DeclareMathAlphabet\Scr{U}{rsf}{m}{n} \DeclareMathAlphabet\mathbi{U}{cmr}{bx}{it}

\def\CY{Calabi-Yau}	
\def\comp{\mbox{\scriptsize \mbox{$\circ \,$}}}

\def\O{\mathcal{O}} \def\c#1{\mathcal{#1}}	
\def\C{{\mathbb C}}\def\P{{\mathbb P}} \def\Q{{\mathbb Q}}\def\RR{{\mathbb R}}\def\F{{\mathbb F}}
 \def\Z{{\mathbb Z}}
\def\D{\mathbf{D}} \def\R{\mathbf{R}}
\def\iso{\cong} 
\def\H{\operatorname{H}}
    
\def\id{\operatorname{id}}
\def\Hom{\operatorname{Hom}} 	
\def\Ext{\operatorname{Ext}}      		\def\RHom{\R\!\operatorname{Hom}}
\def\End{\operatorname{End}}

\def\Spec{\operatorname{Spec}}

	\def\U{\operatorname{U{}}}	

	\def\dim{\operatorname{dim}}

\def\Ltensor{\mathbin{\overset{\mathbf L}\otimes}}
\def\ff#1#2{{\textstyle\dfrac{#1}{#2}}}
		\def\cal{\mathcal}

\def\Cone#1{\operatorname{Cone}\!\left( #1 \right)}
\def\ses#1#2#3{\xymatrix@1{0 \ar[r] & #1 \ar[r] & #2 \ar[r] & #3 \ar[r] & 0}}

\def\card{\operatorname{card}}

\def\pplogo{\vbox{\kern-\headheight\kern -29pt
\halign{##&##\hfil\cr&{\ppnumber}\cr\rule{0pt}{2.5ex}&\ppdate\cr}}}
\makeatletter
\def\ps@firstpage{\ps@empty \def\@oddhead{\hss\pplogo}%
  \let\@evenhead\@oddhead % in case an article starts on a left-hand page
}%      The only change in \maketitle is \thispagestyle{firstpage}  instead of \thispagestyle{plain}
\def\maketitle{\par
 \begingroup
 \def\thefootnote{\fnsymbol{footnote}}
 \def\@makefnmark{\hbox{$^{\@thefnmark}$\hss}}
 \if@twocolumn
 \twocolumn[\@maketitle]
 \else \newpage
 \global\@topnum\z@ \@maketitle \fi\thispagestyle{firstpage}\@thanks
 \endgroup
 \setcounter{footnote}{0}
 \let\maketitle\relax
 \let\@maketitle\relax
 \gdef\@thanks{}\gdef\@author{}\gdef\@title{}\let\thanks\relax}
\makeatother

\def\be{\begin{equation}}
\def\ee{\end{equation}}
\def\calo{{\mathcal O}}
\def\cale{{\mathcal E}}
\def\calA{{\mathcal A}}

\def\calE{{\mathcal E}}
\def\calF{{\mathcal F}}
\def\calH{{\mathcal H}}\newcommand{\roof}[1]{\lceil #1 \rceil}

\begin{document}
\setcounter{page}0
\def\ppnumber{\vbox{\baselineskip14pt
\hbox{hep-th/0605177}}}
\def\ppdate{May 2006} \date{}

\title{\bf \LARGE 
On the geometry of quiver gauge theories\\
{\large (Stacking exceptional collections) }  	\\[20mm]}
\author{{\bf Christopher P.~Herzog} \thanks{herzog@phys.washington.edu}		\\[2mm]
\normalsize  Department of Physics, University of Washington  \\
\normalsize Seattle, WA  98195-1560  USA\\[8mm]
{\bf Robert L.~Karp} \thanks{karp@rutgers.edu}		\\[2mm]
\normalsize  Department of Physics, Rutgers University \\
\normalsize Piscataway, NJ 08854-8019  USA				}

{\hfuzz=10cm\maketitle}

\vskip 1cm

\begin{abstract}
\normalsize
In this paper we advance the program of using exceptional collections to understand the gauge theory description of a D-brane probing a Calabi-Yau singularity. To this end,  we  strengthen the connection between strong exceptional collections and fractional branes.  To demonstrate our ideas,  we derive a strong exceptional collection for every $Y^{p,q}$  singularity, and also prove that this collection is simple.
 \end{abstract}
\vfil\break

\tableofcontents

%%%%%%%%%%%%%%%%%%%%%%%%%%%%%%%%%%%%%%%%%%%%%%%%%%%%%%%%%%%%
\section{Introduction}    \label{s:intro}

In the string theory context, 
fractional branes provide the most straightforward way to determine the low energy gauge theory description of a set of D-branes probing a Calabi-Yau singularity. The simplest case is that of a space-filling D3-brane, but one can consider a vast array of topological charges.  As the D-brane system settles into its lowest energy configuration it usually decays, and the decay products form new bound states. The final configuration of objects is usually referred to as  the fractional branes.

Our main interest in this paper is to determine the fractional branes associated to a given singularity. Since the system preserves supersymmetry, there is an alternate algebraic description.  Let $X$ be a local Calabi-Yau variety.  The fractional branes are a collection of objects $\calA = (A_1, \ldots, A_n)$ in the derived category of coherent sheaves such that
\begin{itemize}
\item $\Ext^0(A_i, A_i) ={\mathbb C}$ \ 
\item $\Ext^q(A_i, A_j) = 0$ for $q\leq 0$ if $i \neq j$ \ 
\end{itemize}
The first condition guarantees that the low energy description is a gauge theory.  The second condition is necessary for stability and supersymmetry. Assuming that the phases of the central charges of 
the $A_i$'s align at the locus in K\"ahler moduli space corresponding to the singularity, the second condition eliminates the tachyons between the $A_i$'s. We will discuss these conditions in more detail in Section~\ref{s:why}.

Exceptional collections provide a simple way of working with and understanding the fractional branes. In many cases they contain only line bundles supported on the exceptional divisor $S$
that partially resolves the Calabi-Yau singularity.

Initially, exceptional collections appeared to have a limited use for understanding
D-brane gauge theories.  Physically relevant collections were known only for del Pezzo surfaces,
which led to the study of Calabi-Yau singularities formed by shrinking a del Pezzo surface, an interesting but restricted class of singularities.
    
More recently, it was discovered that these collections could be used to study a much larger class of singularities \cite{Herzog:2005sy}.\footnote{
We refer to \cite{Herzog:2005sy} for an overview of the vast literature. }
The examples studied were toric, but the method is more general. 
The paper \cite{Herzog:2005sy} left several loose ends, some of which we tie up here.  One underlying problem is that the initial mathematical work on exceptional collections \cite{Rudakov:Book} assumed that $S$ was a smooth variety.  By moving to singular Fano varieties, we need to reinterpret old proofs and occasionally find new ones using the language of smooth stacks.

Another issue is the precise relation between the exceptional collection and the fractional branes.
A central result of \cite{Herzog:2004qw} was that given an exceptional collection
on a del Pezzo $S$ that generated a strong helix, the collection lifts to a set of fractional branes on the total space of the canonical bundle of $S$. In \cite{Herzog:2005sy} we assumed that  the necessary result generalized to the toric singular
cases studied. In this paper, we prove this result using the methods of homological
algebra in Section~ \ref{s:vanish}. In particular, we prove that a {\it full}
exceptional collection on $S$ which generates a strong helix 
lifts to a set of fractional branes ${\mathcal A}$ on the total
space  $KS$ of the canonical bundle $K_S$.  

An important class of examples studied in \cite{Herzog:2005sy} are the $Y^{p,q}$ singularities \cite{Gauntlett:2004yd, Gauntlett:2004zh}, where $p$ and $q$ are both non-negative integers. 
The paper \cite{Herzog:2005sy} provided strong exceptional collections for a fraction of the $Y^{p,q}$'s, in  particular for $Y^{p,p}$, $Y^{p,p-1}$ and $Y^{p,p-2}$.  Here we provide a collection for all $p-q > 2$,
thus completing the list.  We give two proofs that the collection is strong, one in Section~\ref{s:ypq} and one in Section~\ref{s:tor}. The proof in Section~ \ref{s:ypq} is more direct; the actual cohomology groups are calculated explicitly.  The proof in Section \ref{s:tor} on the other hand relies on the Kawamata-Viehweg vanishing theorem. 
This second proof also demonstrates that the associated helix is strong. 

We do not prove that our collection is complete, since this is beyond the scope of the physical applications we have in mind. On the other hand, we have evidence for completeness, and therefore we conjecture that  our collection is complete.

%%%%%%%%%%%%%%%%%%%%%%%%%%%%%%%%%%%%%%%%%%%%%%%%%%%%%%%%%%%%
\section{Preliminary ideas}    \label{s:why}

In this section we present the ideas that underlie the approach taken in the paper. In the first part of the section we give an overview of some of the notions that we need from the theory of quiver algebras,  quiver representations, and their link with exceptional collections on smooth spaces 
\cite{Bondal:FractBranes}. 
Our exposition is partly based on \cite{Herzog:2004qw,Aspinwall:2004vm,Bridgeland:quiver}, to which we also refer for more details and references. In the second part of the section we present a generalization of these ideas using stacks. 

%%%%%%%%%%%%%%%%%%%%%%%%%%%%%%%%%%%%%%%%%%%%%%%%%%%%%%%%%%%%
\subsection{Physical motivation}

Let us start at a point in moduli space where  a D-brane is marginally stable against decaying into a collection of stable constituents, which we call $L_i$. Each constituent may appear with a multiplicity $N_i$, and  is associated a factor of $\U(N_i)$ in the world-volume gauge theory. Since these D-branes are marginally bound, there are massless open strings connecting them, which give chiral fields in the bifundamental representations. This way one associates a quiver gauge theory  to a D-brane decay.

But we need to be more precise about the construction of the quiver. The rules for representing 
topological D-branes as objects in the derived category \cite{Douglas:2000gi} tell us that it is precisely  $\Ext^1(L_i,L_j)$ that counts the number of massless scalars between $L_i$ and $L_j$. The open strings $\Ext^p(L_i,L_j)$ for $p>1$ are massive  and are therefore ignored. On the other hand, the existence of a non-zero $\Ext^0(L_i,L_j)$ would signal a tachyonic instability, indicating that  $L_i$ and $L_j$ formed a bound state, and thus our understanding of this particular decay is incorrect. Therefore the $\Ext^0(L_i,L_j)$'s are forbidden. 

A natural setting where the above described decays happen is when we consider D-branes sitting at singular points of a \CY\ variety. The original context in which quiver gauge theories were introduced is  D-branes  probing a quotient singularity \cite{Douglas:Moore}. The decay products of the 
D0-brane\footnote{It 
  is common practice to mention only the compact dimensions of the brane. Accordingly, a D0-brane  
  in this topological setting could refer to a D3-brane which fills the three non-compact dimensions
  of the full string theory.%
} 
at the singularity are usually referred to as ``fractional branes''. Orbifolds have a vast literature, and we refer to \cite{Paul:TASI2003} for an overview. 

Another class of models is provided by del Pezzo surfaces shrinking down to a point inside a \CY\ variety.\footnote{%
Given a local \CY\ $X$ and a del Pezzo $S$, 
by shrinking down to a point, what we really mean is that there is a partial crepant resolution of the
singularity $\pi: X \to X^*$ where $X^*$ is singular at a point $p$ and $\pi^*(p) = S$.
} 
This case is more general than it appears. One shows that if a  smooth  irreducible divisor in a \CY\ 3-fold is contractible to a point (i.e.~one has a Type~II degeneration) and produces a canonical Gorenstein singularity, then it must be a del Pezzo surface \cite{Wilson:CYKahlercone}.

One can find a set of fractional
branes for the del Pezzo case using the technology of exceptional collections. First recall the following definition:
\begin{definition}
Consider the bounded derived category of coherent sheaves $\D(S)$ on the algebraic variety $S$.
\begin{enumerate}
\item An object $A \in \D(S)$ is called {\em exceptional} if $\Ext^q(A,A) = 0$ for $q\neq 0$ and $\Ext^0(A,A) = {\mathbb C}$.
\item An {\em exceptional collection} $\calA = (A_1, A_2, \ldots, A_n)$ in $\D(S)$ is an ordered collection of exceptional objects such that
\[
\Ext^q(A_i, A_j) = 0 ,\qquad \mbox{for all q, whenever} \; i > j \ .
\]
\item A {\em strong exceptional collection} $\calA$ is an exceptional collection which in addition satisfies: $\Ext^q(A_i, A_j) = 0$ for $q \neq 0$.
\item An exceptional collection is {\em complete} or {\em full} if it generates $\D(S)$.
\end{enumerate}
\end{definition}

The existence of a full and strong exceptional collection for a given variety $S$ constrains the structure of $S$ considerably.  In particular, no smooth projective (and therefore compact) \CY\ variety admits such a collection. The obstruction comes from Serre duality.  The exceptional collections
we are interested in are constructed on an exceptional divisor $S$ of the \CY\ $X$ 
rather than on $X$ itself. 

To simplify matters, and investigate properties inherent to a given del Pezzo singularity, one usually restricts to the neighborhood of the singularity, and calls it a ``local'' \CY\ variety. The local neighborhood of the blow-up is a quasi-projective variety. The adjunction formula and the \CY\ condition tell us that the normal bundle $N_{S/X}$ of the exceptional divisor $S$ equals the canonical bundle of $S$. Therefore, in the vicinity of $S$, the resolved \CY\ looks like the total space of $K_S$. By blowing 
down one shrinks the zero section of $K_S$, and makes the \CY\ locally look like a cone over $S$. Finally one constructs an exceptional collection on $S$. 

The ultimate goal is to understand branes on the ambient local \CY\ $X={\rm Tot} (K_S)$, rather than the subspace $S$. On the other hand, $S$ is embedded in $X$, $\iota\!: S\hookrightarrow X$, and thus there is an induced map $\R \iota_*\!: \D(S)\to \D(X)$. We can think of the objects of $\R \iota_*(\D(S))$ as branes wrapping $S$. Understanding the structure of $\D(S)$ will teach us a great deal about $\D(X)$. As a first step we use a strong full exceptional collection to understand the structure of $\D(S)$.

If $S$ admits a full and strong exceptional collection, then the structure of $\D(S)$ is quite simple, due to a construction that goes back to   Rickard \cite{Rickard:Morita} and Bondal 
\cite{Bondal:Quiver}.\footnote{% 
This construction owes a heavy debt to earlier work by \cite{Beilinson} on projective
space ${\mathbb P}^n$.
}
Namely, if $E_1,\ldots,E_n$ is a full strongly exceptional collection of sheaves on $S$ (or more generally objects in  $\D(S)$) then one  defines the endomorphism algebra
\begin{equation}\label{eq:Aex}
  A = \End(E_1\oplus E_2\oplus\ldots\oplus   E_{n}),
\end{equation}
and Rickard and Bondal prove the following:
\begin{thm}\label{thm:Rickard}
If $S$ has a full strongly exceptional collection $\{E_i\}_{i=1}^n$, then the derived category of coherent sheaves on $S$ is equivalent to the derived category of right $A$-modules $\D(\mbox{\rm mod-}A)$. The equivalence is given by the functor
\begin{equation*}
\Hom_{\D(S)}\left(\bigoplus_{i=1}^n E_i, -\right)\!: \D(S)\longrightarrow \D(\mbox{\rm mod-}A).
\end{equation*}
\end{thm}

Bondal \cite{Bondal:Quiver} observed  in this geometric context 
that the non-commutative  finite-dimensional algebra $A$ can be described as 
the path algebra of a quiver with relations $\c Q$.\footnote{%
The general notion that a finite dimensional algebra can be represented as a quiver is much
older -- see for example \cite{Gabriel}.
}
It is the quiver $\c Q$ that has direct physical relevance, 
with the relations related to the superpotential \cite{KatzAspinwall}. But there is a one-to-one correspondence between left $A$-modules and representations of the quiver $\c Q$ (and we will come to this shortly). Therefore the Rickard-Bondal theorem establishes a correspondence between two very different ways of describing a D-brane: as an object in $\D(S)$  and as a representation of the quiver $\c Q$ \cite{Douglas:Moore,Douglas:Greene:Morrison}. The dictionary goes much deeper, and one shows that the original exceptional objects $E_i$ correspond to certain projective objects in the category of representations of the quiver $\c Q$ (which, from now on, we call $\c Q$-reps). The physical interpretation of the $E_i$'s remains obscure, but they give  convenient generators for $\D(S)$. 

For the definition of the projective counterparts of the $E_i$'s, we need to review a few constructions regarding quivers. This review will also shed some light on the origin of the Rickard-Bondal theorem.

%%%%%%%%%%%%%%%%%%%%%%%%%%%%%%%%%%%%%%%%%%%%%%%%
\subsection{Quiver representations and path algebras}\label{s:why1}

First we recall the definition of the {\em path algebra} of a quiver. Let $\c Q$ be a quiver with nodes $v_i$ and arrows $\alpha$. We consider the paths in $\c Q$, as we flow along the arrows. The nodes are  zero length paths. The path algebra $A$ of $\c Q$ is the $\C$-algebra generated by the paths in the quiver. In particular, every node $v_i$ is associated an element $e_i$, while an arrow $\alpha$ gives an element $a_\alpha$. Multiplication in $A$ is defined by concatenation of paths: 
\begin{equation*}
a_\beta \cdot a_\alpha = \left\{ 
\begin{array}{ll}
a_{\beta \alpha} & \mbox{if $head(\alpha)=tail(\beta)$}	\\
0   & \mbox{otherwise}.
\end{array}	\right.
\end{equation*}
where $\beta \alpha$ is the path consisting of $\alpha$ followed by $\beta$, similarly to function composition. The $e_i$'s are idempotents: $e_i^2=e_i$. 

The path algebra of a {\em quiver with relations} is defined similarly. First we construct the path algebra $A$ as defined above, ignoring the relations. The relations in the quiver mandate certain paths be equal. These equalities generate an ideal of $A$, and we take the quotient $A/I$.

It is natural to look at the category of left $A$-modules $A\mbox{\rm -mod}$. Given 
a left $A$-module $M$, we can construct a 
representation of the quiver as follows. Using 
the idempotent $e_i$, we can form the 
left $A$-module $V_i=e_i M$, which is also  a 
vector space over $\C$. Let $n_i=\dim(V_i)$.  The 
elements of $A$ corresponding to the arrows of $\c Q$ give linear maps between
the $V_i$'s. 
Indeed, for an arrow $\beta$ starting at node $v_i$ and ending at node $v_j$  we have that
\begin{equation*}
 a_\beta \cdot x = e_j \cdot a_\beta \cdot x \in V_j,\qquad \mbox{for any $x\in V_i$.}
\end{equation*}
Therefore we have a set of vector spaces $\C^{n_i}$, one for 
each node, and a set of matrices, one for each arrow. By 
construction, these matrices satisfy the quiver relations, and give a ``quiver representation''. From now on we use the notion of a left $A$-module and that of a $\c Q$-rep interchangeably.

The nodes of a quiver also label  two useful sets of quiver representation. First we have the simple objects. These are the  representations $S _i$ that have no non-trivial subrepresentations. They are particularly easy to describe:  all but the $i$th node is assigned the trivial vector space,  while the $i$th node is assigned the one dimensional vector space $\C$. In a more compact form: $n_j=\delta_{i j}$. All arrows are assigned the $0$ morphism. 

A D0-brane in the quiver language is the representation where every node is assigned the vector space $\C$, while the simple representations are the fractional branes \cite{Douglas:Greene:Morrison,Douglas:2000qw}. In general it is hard to determine  the inverse image of the simple representation $S_i$ under the equivalence of Theorem~\ref{thm:Rickard}. A systematic way of constructing the $S_i$ was presented in \cite{en:fracC2,en:fracC3}, but the examples treated there were all orbifolds.

The second set of quiver representations labeled by the nodes  is defined by $P_i=A e_i$. That is $P_i$ is the subspace of
$A$ generated by all paths {\em starting} at node $i$, and is automatically a left $A$-module, and thus a representation. One shows that the $P_i$'s are {\em projective} objects in the category of $\c Q$-reps.

If the quiver has directed loops, then some of the $P_i$'s may be infinite-dimensional.\footnote{%
 By a directed loop, we mean a path of arrows with the same starting and ending point.
 Sometimes this object is called a cycle.  As our point of view is geometric and cycle has
 other connotations, we prefer loop.
}
The quivers for the del Pezzo surfaces have no such loops, nor do the $Y^{p,q}$ spaces.
%, and it will not be the case in this paper either, for the $Y^{p,q}$ spaces. 
Therefore we assume that there are {\em no\/} directed loops in the quiver $\c Q$
associated to $S$. The quiver associated to $K_S$ in contrast will generically contain
loops.

The simple representations $S_i$ and projective representations $P_i$  are dual in the following sense:
\begin{equation*}
  \Hom(P_i,S_j) = \delta_{ij}\, \C. 
\end{equation*}
The $S_i$'s and $P_i$'s contain all the information encoded in the quiver graph. More precisely, for the $S_i$'s we have:
\begin{equation}\label{b1}
  \dim\Ext^1(S_i,S_j) = n_{ij},\qquad	 \dim\Ext^2(S_i,S_j) = r_{ij},
\end{equation}
where $n_{ij}$ is the number of arrows from node $i$ to node $j$, while $r_{ij}$ is the number of  independent relations imposed on paths from $i$ to $j$.

The  $P_i$'s reconstruct the path algebra $A$ of the quiver directly. To this end one considers the  algebra of endomorphisms $B=\End(\bigoplus_i P_i)$. Multiplication in $B$ is the composition of morphisms. For each $P_i$,  $\End( P_i)=\C$, and we get an idempotent element $f_i$ in $B$. One also notices that $\Hom(P_i,P_j)$ is  the vector space of paths
from $j$ to $i$. A careful analysis shows that 
\begin{equation}\label{e:1}
\End(\bigoplus_i P_i) \iso A^{op},
\end{equation}
where $op$ means that the order of multiplication in   $A$ is reversed:\footnote{This is by definition the opposite algebra. Left modules of  $A$ and right modules  $A^{op}$ are interchanged. This switch
is why one has right modules in Theorem~\ref{thm:Rickard}.}
$a\stackrel{op}{\cdot}b := b\cdot a.$ In light of (\ref{e:1}) the Rickard-Bondal theorem is more natural.

Finally, there is a surprisingly nice relation that links the projective objects to the simple ones, using the technique of mutations. Since we need this notion for both sheaves and quiver representations, we define it in a more general context. Furthermore, mutations are most natural in the context of a derived category. Therefore let $\D(\c A)$ be the derived category of an abelian category $\c A$, and consider two objects $A,B\in \D(\c A)$. The left-mutation of the pair $(A,B)$ is the pair $(\mathsf{L}_{A}B, A)$, where $\mathsf{L}_{A}B$ is defined by\footnote{Our definition differs slightly from that of several authors, like \cite{Aspinwall:2004vm,Rudakov:Book}, but it agrees with others \cite{Karpov:Nogin}, and is more convenient for our purposes.} 
\begin{equation}\label{e:refl}
\mathsf{L}_{A}B=\Cone{ \R\!\Hom_{\D(\c A)}({A},{B})\Ltensor {A}\stackrel{ev}{\longrightarrow }{B} }.
\end{equation}
The map above is the  canonical evaluation morphism. As an immediate consequence of the definition we have the following distinguished triangle
\begin{equation*}
\mathsf{L}_A B[-1] \longrightarrow \RHom(A,B) \Ltensor A \stackrel{ev}{\longrightarrow} B \longrightarrow \mathsf{L}_A B  .
\end{equation*}
We note in passing that left mutations have inverses, called right mutations, and both implement an  action of the braid group on $\D(\c A)$.

Returning to $Q$-reps, let $P_k$ denote the projective $Q$-rep associated to the $k$th node. For any $k\geq 2$ we can define the following ``mutated'' objects: 
\begin{equation}\label{v101}
L_k =  \mathsf{L}_{P_1}\mathsf{L}_{P_2}\ldots\mathsf{L}_{P_{k-1}}P_k.
\end{equation}
As observed by Bondal \cite{Bondal:Quiver, Bondal:FractBranes}, the $L_k$'s defined above are precisely the simple $Q$-reps $S_i$.

%%%%%%%%%%%%%%%%%%%%%%%%%%%%%%%%%%%%%%%%%%%%%%%%
\subsection{$\D(S)$ versus $\D(K S)$}

Now that we have spent some time exploring the structure of $\D(S)$ for a space admitting a full strongly exceptional collection, we can return to examine $\D(K S)$. In the physics literature the transition is usually referred to as the ``completion of the quiver''. Throughout this section we assume that $S$ is smooth.

At this point it is useful to recall a more general  characterization of such spaces:

\begin{thm}[3.1 of \cite{Kawamata:Francia}]\label{thm:KawamataF}
Let $X$ be a variety of dimension $n$ with only log-terminal singularities, and let $K S$ be the total space of the $\Q$-bundle $K_S$. Then
\begin{equation*}
K S = \mbox{\rm Tot}(K_S)= \Spec \left(\bigoplus_{m=0}^\infty \O_S(-m K_S)\right)
\end{equation*}
is an $n + 1$ dimensional variety with only rational Gorenstein singularities and trivial canonical bundle (i.e., $K_{K S}=0$).
\end{thm}

What can we say about $\D(K S)$ given a full strongly exceptional collection on $S$? The simplest case is when 
\begin{equation}\label{e:2}
\dim K_0(S) \otimes \C=\dim S +1. 
\end{equation}
This number -- the rank of the Grothendieck group --  also equals the length of a full exceptional collection. The projective space $\P^n$ is an example satisfying the condition. In general the class of strong exceptional collections is not closed under mutations. On the other hand, Bondal and Polishchuk \cite{Bondal:Polishchuk} introduced the class of  ``geometric'' strong exceptional collections that is. They showed that these collections exist only on varieties satisfying (\ref{e:2}). Obviously, condition  (\ref{e:2}) is extremely restrictive.\footnote{%
A more general formulation of this ``geometric'' condition but
which still restricts to algebras with only quadratic relations was presented in \cite{Hille}.
}

Bridgeland gave an alternative characterization of the geometric condition, assuming that (\ref{e:2}) is satisfied. We modify his definition to fit the general case.
\begin{definition}
An exceptional collection $(E_1,\cdots, E_{n})$ on $S$ is {\em simple}, if for any integers $1\leq i,
j\leq n$ and any $p\geq 0$, for $k>0$
\begin{equation*}
\Ext_{S}^k(E_i\otimes K_S^p,E_j)=0.
\end{equation*} 
\end{definition}
In particular, a simple exceptional collection is  strong.
Given this definition, Bridgeland proves an extension of the Rickard-Bondal theorem \cite{Bridgeland:quiver}.\footnote{Strictly speaking, Bridgeland's original proof of Prop.~4.1 in \cite{Bridgeland:quiver} assumed (\ref{e:2}), but the proof works using our definition of simple without assuming (\ref{e:2}).} 
\begin{prop}\label{Bridgeland:quiver1}
Let $(E_1,\cdots, E_{n})$ be a simple collection on $S$. Denote by $\pi\!: K S \to S$ the projection map of the canonical bundle, and define the algebra $B=\End_{K S} (\bigoplus_{i=1}^{n} \pi^* E_i )$. 
Then the functor 
\begin{equation*}
\Hom_{\D(K S)}\big(\bigoplus_{i=1}^{n} \pi^* E_i, -\big)\colon \D(K S) \longrightarrow \D(\mbox{\rm mod-} B)
\end{equation*}
is an equivalence of categories.
 \end{prop}
The algebra $B$  is infinite-dimensional, but Bridgeland shows that $B$ is the path algebra of a quiver with relations $\c Q_B$. The algebra $A$, as defined in (\ref{eq:Aex}), is a subalgebra of  $B$. To see this, first recall that $\pi^*$ and $\pi_*$ are an adjoint pair:
\begin{equation}\label{3}
\Hom_{K S}(\pi^*E,\pi^* F)=\Hom_{ S}(E,\pi_*\pi^* F).
\end{equation} 
The projection formula gives $\pi_*\pi^* F=F\otimes \pi_* (\O_{K S})$.  Theorem~\ref{thm:KawamataF} implies that 
\begin{equation}\label{4}
\pi_* (\O_{K S})=\bigoplus_{p\leq 0} K_S^p = \O_S\oplus\bigoplus_{r\leq -1} K_S^r.
\end{equation}
Now 
\begin{equation*}
B=\End_{\D(K S)} \big(\bigoplus_{i=1}^{n} \pi^* E_i\big)=
\bigoplus_{i=1}^{n}\bigoplus_{j=1}^{n}\Hom_{K S}\big(\pi^* E_i,\pi^* E_j\big).
\end{equation*} 
Using (\ref{3}) and (\ref{4}) we obtain
\begin{equation*}
B=\bigoplus_{i=1}^{n}\bigoplus_{j=1}^{n}\Hom_{ S} \big( E_i, E_j\big)\oplus \cdots =A\oplus \cdots.
\end{equation*} 
It is the quiver associated to $B$ that is the completed quiver of physical interest. The obvious question to ask at this point is how do we construct this quiver?

The simplest example to consider is $K \P^n$. $\P^n$ has the well-known {\em geometric} exceptional collection of length ${n+1}$
\begin{equation}\label{5}
\O,\O(1),\ldots ,\O(n). 
\end{equation}
The quiver describing the $A$ algebra is the Beilinson quiver: 
\begin{equation*}
\xymatrix@1@=1.6cm{\bullet \ar[r]^-{n+1}& 	 \bullet \ar[r]^-{n+1}& 	*{\bullet \quad\cdots\quad \bullet} \ar[r]^-{n+1}& \bullet }
\end{equation*} 
The quiver for the algebra $B$  turns out to be the McKay quiver of $\C^{n+1}/\Z_{n+1}$, where the $\Z_{n+1}$ action has weights $(1\ldots ,1)$, i.e., it is supersymmetric. The appearance of the McKay quiver in this context is not accidental. Rather it is a manifestation of the fact that the resolution of the $\C^{n+1}/\Z_{n+1}$ singularity can be thought of as  $K \P^n$, and thus in this case Bridgeland's Prop.~\ref{Bridgeland:quiver1} is a consequence of the McKay  correspondence.

%%%%%%%%%%%%%%%%%%%%%%%%%%%%%%%%%%%%%%%%%%%%%%%%
\subsection{The physical quiver}

Using the algebra $B$ to obtain the physical quiver is equivalent to using the projective objects in the category of $\c Q_B$-reps. But even in this simple case, it is not completely elementary to write down the extended quiver $\c Q_B$. On the other hand we saw in Section~\ref{s:why1} that the simple objects contain the same information. So how do we go from the simples of the quiver to the simples of the completed quiver? Our approach is to use the geometric representation of the simples. As it turns out this gives a more straightforward procedure to obtain the quiver.

Let us assume that we determined the dual collection $S_i$ in terms of sheaves (complexes), for example by using mutations as explained at the end of Sec.~\ref{s:why1}. For $\P^n$ the mutation theoretic dual of the exceptional collection (\ref{5}) is
\begin{equation}\label{7}
\Omega^{n}(n)[n],\ldots,\Omega^{1}(1)[1],\O.
\end{equation} 
Since $K S$ is the total space of the line bundle $K_S$, the zero section of $K_S$ embeds $S$ into $K S$. We call the embedding $\iota$:
\begin{equation*}
\iota\colon S\hookrightarrow K S.
\end{equation*} 
Using $\iota$ we can make the following observation:\footnote{We are not aware of the existence of this fact in the literature.}

\begin{lemma}
Given a dual pair $\{P_i\}$ and $\{S_j\}$ on $S$, the pair $\{\pi^* P_i\}$ and $\{\iota_* S_j\}$ is a dual pair on $K S$.
\end{lemma}

\begin{proof}
We need to compute $\Hom_{K S}(\pi^* P_i, \iota_* S_j)$. For this we recall that $\pi_*$ is the right adjoint of $\pi^*$, and thus
\begin{equation*}
\Hom_{K S}(\pi^* P_i, \iota_* S_j) = \Hom_{S}(P_i, \pi_* \iota_* S_j).
\end{equation*} 
But $\pi_* \iota_* = (\pi\comp \iota)_*$, while $\pi \comp\iota=\id$ since $\iota$ is a section. The claim then follows from the fact that $\{P_i\}$ and $\{S_j\}$ is a  dual pair.
\end{proof}

The lemma implies that the $\iota_* S_j$'s are the simples of the extended quiver $\c Q_B$. Using (\ref{b1}), the arrows of $\c Q_B$ are given by the $\Ext^1$'s between the $\iota_* S_j$'s. To compute these $\Ext^1$'s on $K S$ we can use a spectral sequence which is a consequence of the local to global spectral sequence. Namely, if $\iota \!: S \hookrightarrow X$ is an embedding, and $N_{S/X}$ the normal bundle of $S$ in $X$, then there is  a spectral sequence with $E_2$ term:
\begin{equation}\label{SS1}
E_2^{p,q}=
\Ext^p_S( {\cal E} ,\, {\cal F} \otimes \Lambda^q N_{S/X}) \: \Longrightarrow \:
{\Ext}^{p+q}_X\left( \iota_* {\cal E}, \iota_* {\cal F} \right)
\end{equation}
where $\Lambda^q$ denotes the $q$th exterior power. 

In our case $N_{S/K S}=K_S$, and using Serre duality we obtain
\begin{equation*}
\Ext_{K S}^q(\iota_* S_i, \iota_* S_j) = \Ext_{S}^q(S_i,  S_j)\oplus \Ext_{S}^{d+1-q}(S_j,  S_i)^\vee ,
\label{pushforward}
\end{equation*} 
where $d$ is the dimension of $S$.

Using the  spectral sequence (\ref{SS1}) it is easy to show that the $\Ext^1$-quiver of the collection (\ref{7}) is indeed the McKay quiver:
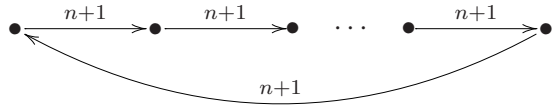
\begin{equation*}
\xymatrix@1@=1.6cm{\bullet \ar[r]^-{n+1}& 	 \bullet \ar[r]^-{n+1}& 	*{\bullet \quad\cdots\quad \bullet} \ar[r]^-{n+1}
& \bullet  \ar@/^1cm/[lll]_{n+1}}
\end{equation*} 

From a physical point of view we are describing the decay of a D0-brane in $K S$ sitting inside $S$. When $S$ shrinks and the D0 gets destabilized, we expect the physics to localize in the neighborhood of $S$, and that only states supported on this neighborhood would have normalizable wavefunctions. The exceptional collection on $S$ knows nothing about the neighborhood of $S$. The extra data can be prescribed by the normal bundle $ N_{S/K S}$ of $S$ in $K S$. A priori it is not clear how to use $ N_{S/K S}$ to ``fix'' the exceptional collection on $S$. Using the dual collection, i.e., the simples $S_i$, the ``fix'' is straightforward. The $\iota_* S_i$'s are the fractional branes.\footnote{Katz and Sharpe \cite{Eric:DC1} gave a direct physical interpretation to the spectral sequence (\ref{SS1}) using vertex operators, explaining the physical meaning of $\iota_*$.} The $\Ext^1$ quiver $\c Q_B$ of $\iota_* S_i$ is the quiver describing the decay.

%If one is interested only in the quiver and not the decay products, then the method of Higgsing 
%\cite{Morrison:Plesser} is applicable. One finds that for del Pezzos Higgsing \cite{} and exceptional
% collections \cite{} give the same answer. This will be the case in the present paper as well. 

Our approach of constructing $\c Q_B$ can be summarized in the following four steps:
\begin{enumerate}
\item construct a  full and simple exceptional collection on $S$,
\item construct the dual collection $S_i$,
\item using the zero section $\iota\colon S\hookrightarrow K S$, construct $\iota_* S_i$,
\item compute the $\Ext^1$-quiver of the $\iota_* S_i$'s.
\end{enumerate}

%%%%%%%%%%%%%%%%%%%%%%%%%%%%%%%%%%%%%%%%%%%%%%%%
\subsection{The stacky version}	\label{sec:stacky}

So far we presented a way to understand D-brane decay at the singularity in $K S={\rm Tot}(K_S)$ created by contracting the zero section. Throughout $S$ was assumed to be smooth. Unfortunately, our main  examples in this paper, the $Y^{p,q}$ spaces, have their base $S$ singular. But as we will see shortly, they have only quotient singularities.

It is natural to ask whether any of the structure outlined earlier survives if one has either a singular divisor, or a reducible divisor and one only partially resolves the space, and therefore ends up with a singular \CY\ geometry, or both. On the negative side, one knows that the derived category of coherent sheaves on a singular space is ill behaved, and in particular it is not equivalent to the derived category on the resolved smooth space \cite{TomBridgeland:discussion}. But there is a well-known way of ``beautifying'' singular schemes, using stacks. In the language of stacks, quotient singularities become smooth, and the hope for an equivalence is revived. To this aim we cite a recent result of Kawamata:
\begin{thm}[4.5 of \cite{Kawamata:GMcKay}]\label{thm:Kawamata}
Let $X$ and $Y$ be projective toric varieties with only quotient singularities, and let $\c X$ and $\c Y$ be the associated smooth Deligne-Mumford stacks. Let $f\colon X\to Y$ be a toric proper birational map which is crepant in the sense that $g^*(K_X ) = h^*(K_Y  )$ for two toric proper birational morphisms from a common toric variety $Z$ that makes the diagram
\begin{equation*}
\xymatrix{
  &Z\ar[dl]_{g}\ar[dr]^{h}&\\
  X \ar[rr]_{f}& & Y\,.}
\end{equation*}
commute. Then there is an equivalence of triangulated categories $F\! : \D(\c Y) \to \D(\c X)$.
\end{thm}

Applied to the case of a toric\footnote{
The theorem is conjectured to hold for non-toric varieties, see, e.g., Sec.~2 of \cite{Kawamata:GMcKay}.} 
partial {\em crepant} resolution $f\! : \tilde{X}\to X$, the theorem implies that  there is an equivalence of triangulated categories between $\D(\c {\tilde{X}})$ and $\D(\c X)$. But the theorem also implies that 
\begin{equation}\label{v1}
\D(\c X)\cong \D(\mbox{{\bf partial} crepant  resolution})\cong \D(\mbox{crepant resolution}).
\end{equation}
In particular, for $G$ abelian it proves the derived McKay correspondence \cite{Mukai:McKay}:
\begin{equation*}
\D([\C^n/G])\cong \D(\mbox{crepant resolution of $\C^n/G$}).
\end{equation*}

The equivalence in (\ref{v1}) gives a flexible framework to work in. What we want is D0 decay in the singular space   $X$. In the previous sections we worked in the totally resolved space, and there were no partial resolutions. In light of Theorem~\ref{thm:Kawamata} we can work in a partially resolved space as well. Since we work with topological B-branes, and the resolution is a K\"ahler deformation, all these models are equivalent. This is in fact the content of Theorem~\ref{thm:Kawamata}.

The other ingredient used in the first part of this section is the Rickard-Bondal theorem \ref{thm:Rickard} providing an  equivalence between the derived category of sheaves with a derived category of modules. Fortunately this one also extends to the stacky case, as proved by Kawamata (Prop.~4.7 in \cite{Kawamata:GMcKay}). Therefore we have all the pieces in place, and we can use the strategy from the end of the previous section.

It is important to note that in the realm of singular varieties as opposed to stacks, the analog of the Rickard-Bondal theorem \ref{thm:Rickard} (or Kawamata's theorem \ref{thm:Kawamata}) is guaranteed to fail. Therefore there is no meaningful identification between the sheafy model of D-branes and the quivery model. 

%%%%%%%%%%%%%%%%%%%%%%%%%%%%%%%%%%%%%%%%%%%%%%%%%%%%%%%%%%%%%%%%
\section{Vanishing results -- absence of tachyons}    \label{s:vanish}

In this section we lay out a set of sufficient conditions, which, if satisfied by an exceptional collection on $S$, lead to a  tachyon free spectrum of fractional branes on the Calabi-Yau $KS$. From (\ref{pushforward}), in order for there to be no tachyons between the fractional branes $\imath_* S_i$ we need the vanishing of the $\Ext^0$ and $\Ext^{d+1}$ groups between the $S_i$'s.
 
In the next subsection we find that the $\Ext^0$ automatically vanish if the $S_i$'s came from a strongly exceptional collection.
However,  the vanishing of the $\Ext^{d+1}$'s  is  more subtle. In subsection \ref{sec:nodp1s} we show that a {\em simple} and {\em full} exceptional collection guarantees that the $\Ext^{d+1}$'s vanish.  

The arguments of this section use only formal properties of a triangulated category of sheaves, and therefore apply to either a smooth variety or a smooth stack. Throughout the section we work on such a smooth space $X$ of dimension $d$.

%%%%%%%%%%%%%%%%%%%%%%%%%%%%%%%%%%%%%%%%%%%%%%%%%%%%%%%%%%%%%%
\subsection{Absence of Ext$^0$'s}

We begin by recalling a standard result about mutations:
\begin{lemma}[Sect. 7.2.3 of \cite{Rudakov:Book}]\label{leftreduce}
Given an exceptional collection $(A,B,C)$, then for any $q\in\Z$
\begin{equation*}
\Ext^q(\mathsf{L}_A B, \mathsf{L}_A C) = \Ext^q(B,C) \ .
\end{equation*} 
\end{lemma}

Next we  prove two auxiliary results that we will need in the sequel.
\begin{lemma}\label{itoiminusone}
Given an exceptional pair $(A,B)$ in $\D(X)$, then for any $q\in\Z$
 \[
\Ext^{q+1}(\mathsf{L}_A B, A) = \Ext^{-q}(A,B) \ .
\]
\end{lemma}
\begin{proof}
Applying  the contravariant functor $\Hom(\cdot, A)$ to the defining equation of left mutation, we find the long exact sequence (LES)
\[
\cdots \to \Ext^q(\mathsf{L}_A B, A) \to \Ext^q(B,A) \to \Ext^q(\RHom(A,B)\otimes A, A) \to \cdots
\]
$\Ext^q(B,A) = 0$ for all $q$ since $(A,B)$ forms an exceptional pair. Furthermore, one can show that $\Ext^q(\RHom(A,B) \otimes A, A) = \Ext^{-q}(A,B)$ due to the fact that $A$ is an exceptional object. The claim then follows immediately.
\end{proof}

\begin{lemma}\label{lowerbound}
Given an exceptional collection $(A,B,C)$ such that $\Ext^q(A,B) = 0$ for $q \neq 0$ and $\Ext^{q}(A,C) = \Ext^{q+1}(B,C) = 0$ for $q \geq m$, then $\Ext^{q}(A, \mathsf{L}_B C) = 0$ for $q \geq m$.
\end{lemma}
\begin{proof}
Consider the covariant functor $\Hom(A, \cdot)$.  Applying this functor to the definition of left mutation, we find the LES
\[
\cdots \to \Ext^q(A, \RHom(B,C) \otimes B) \to \Ext^q(A,C) \to \Ext^q(A, \mathsf{L}_B C) \to \cdots
\]
Because of the constraint on $\Ext^q(A,B)$, it follows that
$\Ext^q(A, \RHom(B,C) \otimes B) = \Ext^q(B,C)\otimes \Ext^0(A, B)$. 
The claim then follows from the LES and the constraints on $\Ext^q(A,C)$ and $\Ext^q(B,C)$.
\end{proof}

Given an exceptional collection $\calE = (E_1, E_2, \ldots, E_n)$, in light of Eq.~(\ref{v101}) we define the dual collection to be
$\calF = (F_n, F_{n-1}, \ldots, F_1)$, where 
\begin{equation*}
F_j = \mathsf{L}_{E_1} \mathsf{L}_{E_2} \cdots \mathsf{L}_{E_{j-1}} E_j  .
\end{equation*} 
These $F_i$'s are the geometric counterparts of 
the simple quiver representations $S_i$ of Section \ref{s:why}. 

\begin{prop}\label{pr1}
The dual collection of an exceptional collection is also exceptional.
\end{prop}
\begin{proof}
Prop. 8.2.1 (pg. 77 of \cite{Rudakov:Book}) shows that given  an exceptional collection $\calE= (E_1, E_2, \ldots, E_n)$, the collection
$$\calE' = (E_1, \ldots, \mathsf{L}_{E_{j-1}} E_j, E_{j-1}, E_{j+1}, \ldots, E_n)$$
obtained by mutating $E_j$ over its neighbor $E_{j-1}$ is also exceptional. Since the definition of the dual collection involved a succession of left mutations, the end result is exceptional as well.
\end{proof}

Now we are ready to prove  the absence of tachyons coming from $\Ext^0$.
\begin{prop}\label{firstvanishing}
Let $\calE$ be a strong exceptional collection and $\calF$ its dual collection.  For two different elements $F_i$ and $F_j$ of $\calF$, we have $\Ext^q(F_i, F_j) = 0$ for $q \leq 0$.
\end{prop}
\begin{proof}
For $i<j$, the result follows from Prop.~\ref{pr1}.  Therefore we focus on the case $i>j$.  By definition
\[
\Ext^q(F_i, F_j) = \Ext^q(\mathsf{L}_{E_1} \cdots \mathsf{L}_{E_{i-1}} E_i, \mathsf{L}_{E_1}\cdots \mathsf{L}_{E_{j-1}} E_j) \ .
\]
Using Lemma \ref{leftreduce} repeatedly gives
\[
 \Ext^q(\mathsf{L}_{E_1} \cdots \mathsf{L}_{E_{i-1}} E_i, \mathsf{L}_{E_1}\cdots \mathsf{L}_{E_{j-1}} E_j) =
 \Ext^q(\mathsf{L}_{E_j} \cdots \mathsf{L}_{E_{i-1}} E_i, E_j) \ .
\]
From Lemma~\ref{itoiminusone} we have
\[
\Ext^q(\mathsf{L}_{E_j} \cdots \mathsf{L}_{E_{i-1}}E_i, E_j) = \Ext^{-q+1} (E_j, \mathsf{L}_{E_{j+1}} \cdots \mathsf{L}_{E_{i-1}} E_i) \ .
\]
Consider the vector spaces
\[
M^q(j,i,k) \equiv \Ext^q(E_j, \mathsf{L}_{E_i} \mathsf{L}_{E_{i+1}} \cdots \mathsf{L}_{E_{i+k-1}} E_{i+k})
\]
 for 
$1 \leq j<i \leq i+k\leq n$.
The rest of the proof follows by induction on $k$. We will show that $M^q(j,i,k) = 0$ for $q\geq 1$ and any $j$, $i$, and $k$ in the allowed range.  

Consider first the case $k=0$. Then $M^q(j,i,0) = \Ext^q(E_j, E_{i})$, and this vanishes for $q\neq0$ because the initial collection was strongly exceptional.  Thus $M^q(j,i,0)=0$ for $q \geq 1$.

Now assume that $M^q(j,i,k)=0$ for $q \geq 1$ and any $i>j$; consider $M^k(j,i,k+1)$.   
We can apply Lemma \ref{lowerbound}, setting $A=E_j$, $B=E_i$,  
$C = \mathsf{L}_{E_{i+1}} \cdots \mathsf{L}_{E_{i+k}} E_{i+k+1} $ and  $m=1$. $\Ext^q(A,B)=0$ for $q\neq 0$ since $\cale$ is strongly exceptional. While the remaining conditions of the Lemma are met by the inductive assumptions on $M^q(i, i+1,k)$ and $M^q(j,i+1,k)$. The Lemma then implies that $M^q(j,i,k+1)=0$ for $q\geq 1$.
\end{proof}

%%%%%%%%%%%%%%%%%%%%%%%%%%%%%%%%%%%%%%%%%%%%%%%%%%%%%%%%%%%%%%
\subsection{Absence of Ext$^{d+1}$'s} \label{sec:nodp1s}

In this part, we prove  that if the exceptional collection is {\em simple} and {\em full} 
then the $\Ext^{d+1}$'s vanish as well. We actually prove a slightly stronger result, replacing the 
simple assumption with the assumption of a strong helix \cite{Herzog:2005sy} which we define below.
This result completes the proof of the tachyon freeness for the fractional branes $\imath_* S_i$.

Once again, let $X$ be a smooth variety or  stack of dimension $d$, with canonical bundle $K$.
A {\em helix of period $n$} is a bi-infinite extension $\{E_i\}_{i\in \Z}$ of an exceptional collection $\c E=(E_1,\cdots ,E_n)$ satisfying
\begin{equation*}
E_{i-n}\iso E_i\otimes K,\quad\text{ for all }i\in \Z.
\end{equation*} 

Helices are closely related to left mutations. Bondal proves (Theorem 8.4.1 of \cite{Rudakov:Book}) that for a full exceptional collection $\cale$ 
\begin{equation}\label{l:x1}
\mathsf{L}_{E_1} \mathsf{L}_{E_2} \cdots \mathsf{L}_{E_{n-1}} E_n = E_n \otimes K[d].
\end{equation} 
Strictly speaking, the original proof is presented for smooth varieties, but it directly extends to smooth stacks.

We are only interested in helices generated by full exceptional collections. A {\em foundation of a helix} is a minimal, finite set of consecutive elements of a helix which form a full exceptional collection. Note that if there exists a foundation of length $n$, then by Serre duality and Lemma~8.2.2 of \cite{Rudakov:Book} (which states that a mutation of a foundation is again a foundation), any set of $n$ consecutive elements is also a foundation. This result motivates the following definition:
\begin{definition}
A {\em strong helix} is a helix where any foundation is a strong exceptional collection.
\end{definition}
Bridgeland's simple criterion for a full exceptional collection implies that the helix
must be strong.  However, the converse is not necessarily true -- a strong helix only guarantees
the simple criterion for $p=0$ and $p=1$.  

\begin{prop}
Let $\calH$ be a strong helix on $X$, and choose a foundation $\calE$ of $\calH$.  Let $\calF$ be the dual of $\calE$.  Then for any two elements $F_i, F_j \in \calF$, $\Ext^q(F_i, F_j) = 0$ for $q \geq d+1$.
\end{prop}
\begin{proof}
Write $\calF = (F_n, \ldots, F_1)$. Then Prop.~\ref{pr1} guarantees that $\c F$ is an exceptional collection, and thus $\Ext^q(F_i, F_j) = 0$ for $i\leq j$ and $q>0$. So we need to consider only the case $i>j$.

Since  $\calH$ is a strong helix, the adjacent foundation $\calE' = (E_n \otimes K, E_1, E_2, \ldots, E_{n-1})$ is a strong exceptional collection. The dual of $\calE'$ is
\begin{eqnarray*}
\calF' &=& (F_n', F_{n-1}', \ldots, F_1') \\
&=& 
(\mathsf{L}_{E_n \otimes K} F_{n-1}, \mathsf{L}_{E_n \otimes K} F_{n-2}, \ldots, \mathsf{L}_{E_n \otimes K} E_1, E_n \otimes K) \ .
\end{eqnarray*}

On the other hand, for $j> 1$, 
\begin{eqnarray*}
\Ext^q(F_j', F_1') &=& \Ext^q(\mathsf{L}_{E_n \otimes K} F_{j-1}, E_n \otimes K) \\
&=& \Ext^{-q+1}(E_n \otimes K, F_{j-1}) \qquad\mbox{by Lemma \ref{itoiminusone}}\\
&=& \Ext^{-q+d+1}(E_n \otimes K[d], F_{j-1})\\
&=& \Ext^{-q+d+1}(F_n, F_{j-1})  \qquad\mbox{by Eq. (\ref{l:x1})}\ .
\end{eqnarray*}
By Prop. \ref{firstvanishing}, $\Ext^q(F_j', F_1') = 0$ for $q\leq 0$, and thus $\Ext^q(F_n, F_{j-1}) = 0$ for $q \geq d+1$.

Since we started with a strong helix, we can repeat the argument with $\calE'$ playing the role of $\calE$, and the above argument shows that $\Ext^q(F_n', F_{j-1}') = 0$ for $q \geq d+1$ and $j> 1$. But for $j> 2$,
\begin{eqnarray*}
\Ext^q(F_n', F_{j-1}') &=& \Ext^q(\mathsf{L}_{E_n\otimes K} F_{n-1},
\mathsf{L}_{E_n \otimes K} F_{j-2}) \\
&=&  \Ext^q(F_{n-1}, F_{j-2}) \qquad\mbox{by Lemma \ref{leftreduce}}
\end{eqnarray*}
Thus $\Ext^q(F_{n-1}, F_{j-2}) = 0$ for $q \geq d+1$ and $j> 2$. The full result then follows by induction, showing that $\Ext^q(F_{n-k}, F_{j-1-k}) = 0$ for $q \geq d+1$, $0\leq k<n$, and $j> k+1$.
\end{proof}

%%%%%%%%%%%%%%%%%%%%%%%%%%%%%%%%%%%%%%%%%%%%%%%%%%%%%%%%%%%%%
\subsection{Invariance of the quiver}

Given a complete exceptional collection $\calE$, we generated a helix $\calH$ through tensoring by $K$.  In this section, given an arbitrary foundation $\calE'$ of $\calH$, we try to understand its dual, $\calF'$, in terms of $\calF$, the dual of $\calE$.

Let $\calE = (E_1, \ldots, E_n)$ and let the neighboring foundation be $\calE' = (F_n[-d], E_1, \ldots, E_{n-1})$.  Then the dual collection takes the form
\be
\calF' = (\mathsf{L}_{F_n[-d]} F_{n-1}, \ldots, \mathsf{L}_{F_n[-d]} F_1, F_n[-d]) \ .
\ee

Consider the $\Ext$'s of $\calF'$.  These are identical to those of $\calF$ except for pairs involving $F_n[-d]$: 
\begin{equation*}
\Ext^q(\mathsf{L}_{F_n[-d]} F_j, F_n[-d]) = \Ext^{-q+1}(F_n[-d], F_j) = \Ext^{d+1-q}(F_n, F_j)  .
\end{equation*}
This relation
exchanges the $\Ext^1$'s of $F_n$ to $F_j$ with the $\Ext^{d}$'s of $\mathsf{L}_{F_n[-d]} F_j$ to $F_n[-d]$, and vice versa. 
The result (\ref{pushforward}) then implies that the extended quiver associated to $\calF$ is the same as the one associated to $\calF'$.

%%%%%%%%%%%%%%%%%%%%%%%%%%%%%%%%%%%%%%%%%%%%%%%%%%%%%%%%%%%%%%%%
\section{The geometry of the $Y^{p,q}$ spaces}    \label{s:ypq}

The $Y^{p,q}$ spaces, for integers $0 \leq q \leq p$, are three dimensional toric 
varieties.\footnote{The case $Y^{0,0}$ is excluded.}
 They appeared first in the physics literature as a set of Ricci flat
metrics \cite{Gauntlett:2004yd, Gauntlett:2004zh}.  Later,
\cite{Martelli:2004wu} classified these three-folds using toric
geometry.  At last, \cite{Benvenuti} provided gauge theories on these
spaces using physics techniques. Our aim is to obtain a complementary
understanding of these  gauge theories, using the mathematical method
of exceptional collections on a toric stack.
  
We begin by investigating the geometry of the $Y^{p,q}$ spaces as stacks.
The toric fan consists of one cone generated by the following four vectors
\begin{equation}	\label{e:ypqpts}
V_1 = (1,0,0)\ , \; \; V_2 = (1, 1,0)\ ,\; \; V_3 = (1, p,p)\ , \; \; V_4 = (1,p-q-1,p-q).
\end{equation} 

In \cite{Herzog:2005sy} we provided strong exceptional collections on several $Y^{p,q}$ spaces.  In particular, we treated $Y^{p,p}$, which are orbifolds of ${\mathbb C}^3$,\footnote{In this case the fractional branes can be constructed quite explicitly \cite{en:fracC3}.} $Y^{p,p-1}$ and $Y^{p,p-2r}$, for  $\gcd(p,r)=1$  \cite{Herzog:2005sy}.  In this paper, we will give a strong exceptional collection for all  $p-q-2 \geq 0$.  For $q=p-2r$ this new collection is different from the one in \cite{Herzog:2005sy}.

We can partially resolve the $Y^{p,q}$ space by blowing up the various toric divisors. If we choose the points in the plane $x=1$ in the interior of the polygon generated by $V_1,\ldots ,V_4$ then the partial resolution is crepant. In this paper we choose to  blow up $V=(1,1,1)$, and we denote the partially resolved space $\tilde{Y}^{p,q}$. 

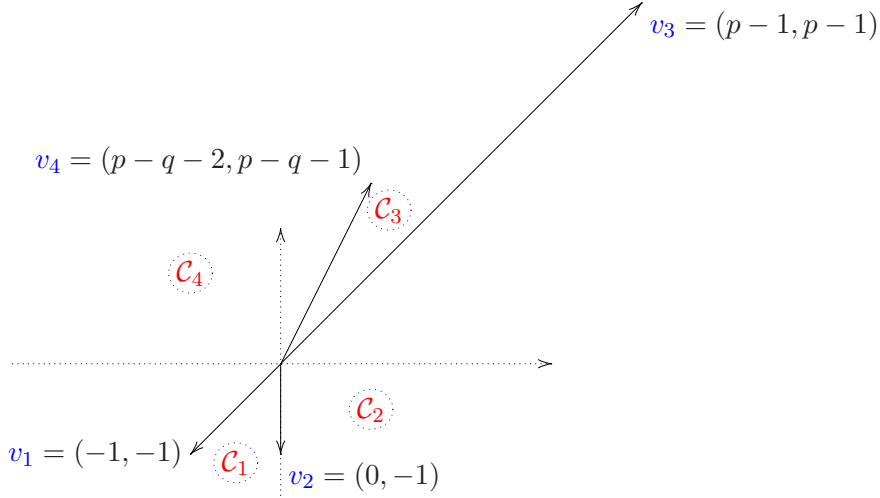
\begin{figure}[h] 
\begin{equation}\nonumber
\begin{xy} <1.2cm,0cm>:
 {\ar (0,0);(-1,-1) *+!R{{\color{blue}v_1}=(-1,-1)}}
,{\ar (0,0);(4,4)    *+!LU{{\color{blue}v_3}=(p-1,p-1)}}
,{\ar (0,0);(0,-1)  *+!LU{{\color{blue}v_2}=(0,-1)}}
,{\ar (0,0);(1,2)    *+!RD{{\color{blue}v_4}=(p-q-2,p-q-1)}}
,{\ar@{-}@{.>} (0,-1.5);(0,1.5) }
,{\ar@{-}@{.>} (-3,0);(3,0) }
,(-.5,-1.1)*+[o][F.]{\color{red} {\mathcal C}_1}
,(1.2,1.7)*+[o][F.]{\color{red} {\mathcal C}_3}
,(1,-.51)*+[o][F.]{\color{red} {\mathcal C}_2}
,(-1,1)*+[o][F.]{\color{red} {\mathcal C}_4}
\end{xy}
\end{equation}
  \caption{The toric fan for the $X_{p,q}$ space.}
  \label{f:fan}
\end{figure}

As usual in toric geometry, the  divisor $D_V$ corresponding to $V=(1,1,1)$ is also a toric variety. We call it $X_{p,q}$. The toric fan of $X_{p,q}$ is obtained from (\ref{e:ypqpts}), and is depicted in Fig.~\ref{f:fan}. 

The rays in question are
\begin{equation}\label{e:weights}
\begin{split}
&v_1 = (-1,-1), \; \;  v_2 = (0,-1), \\
&v_3 = (p-1,p-1), \; \; v_4 = (p-q-2,p-q-1) \, .
\end{split}
\end{equation}

By construction $\tilde{Y}^{p,q}$ is the total space of the canonical sheaf over the toric surface $X_{p,q}$. Thus $\tilde{Y}^{p,q}$ is an example of a space discussed in Sec.~\ref{s:why}. The linear equivalence relations among the toric divisors of $X_{p,q}$ are
\begin{equation}\label{x5}
D_4 \sim D_2,\quad D_1 \sim (p-1)D_3+(p-q-2)D_4.
\end{equation}

We can gain  insight into the structure of this space by using the stacky modification \cite{BorisovDM} of Cox's holomorphic quotient construction \cite{Cox:HoloQout}. Let $x_1,\ldots , x_4$ be coordinates on $\C^4$. Then $X_{p,q}$ is the quotient of $\C^4-\{x_1=x_3=0,x_2=x_4=0\}$ by $\C^*\times \C^*$, where the weights of the two $\C^*$ actions can be read out from (\ref{e:weights}): 
\begin{equation}\label{c1}
(\lambda_1^{p-1}  \lambda_2^{p-q-2} x_1, \lambda_2 x_2, \lambda_1 x_3, \lambda_2 x_4)\,, 
\end{equation}
and $( \lambda_1 , \lambda_2)\in \C^*\times \C^*$.
 
From the holomorphic quotient perspective, 
consider first the $x_1=0$ subspace; call it $\c D_1$. Since the subset $\{x_1=x_3=0\}$ is excluded and  $x_1=0$ we must have $x_3\neq 0$. Therefore we can use the first $\C^*$ action to completely fix $x_3$, say to $x_3=1$. What we are left with is $\C^2-\{x_2=x_4=0\}/\C^*\iso \P^1$. 

The same answer is obtained if we use the stacky method of \cite{BorisovDM}.\footnote{An easily readable account of some of the results in \cite{BorisovDM} is given in the Appendix of \cite{Pantev:2005zs}.} The lattice associated to $\c D_1$ is the quotient
of $\Z^2$ by the subgroup generated by $v_1=(-1,-1)$:
$N_{\c D_1}= \Z^2/\langle v_1\rangle$. 
%In other words, it is the quotient of $\Z^2$ by the subgroup generated by $v_1=(-1,-1)$. 

More generally, if we consider the abelian group $\Z^2$, and the subgroup $\langle (a,b)\rangle$ generated by $(a,b)$, for $a,b \in \Z$, then the quotient $\Z^2/\langle (a,b) \rangle$ is again a finitely generated abelian group. By the fundamental theorem of finitely generated abelian groups it is necessarily of the form $\Z^r\oplus \mbox{finite torsion}$. In fact it is easy to show that $\Z^2/\langle (a,b)\rangle\iso \Z\oplus \Z_{\gcd(a,b)}$. 

From this observation, it is automatic that $N_{\c D_1}\iso \Z$. We choose the following isomorphism: first we do a change of basis in $\Z^2$, with  new  basis vectors $(-1,-1)$ and $(0,1)$. Then $\psi\colon N_{\c D_1}=\Z^2/\langle (-1,-1) \rangle \to \Z$ is projection on the subspace generated by $(0,1)$. The $Star$ of $\c D_1$ consists of the cones $\c C_1$ and $\c C_4$. The image of $v_2=(0,-1)$ in $N_{\c D_1}\iso \Z$ is $-1$. Similarly, the image of $v_4=(p-q-2,p-q-1)$ in $N_{\c D_1}\iso \Z$ is $1$. Running the BCS construction this gives a $\P^1$, as expected. 

The $x_2=0$ and $x_4=0$ subspaces are  also straightforward. Consider the subspace $\c D_2$ given by $x_2=0$. Once again, the $\{x_2=x_4=0\}$ subset is excluded, and the second $\C^*$ action  completely fixes $x_4$, yielding $\C^2-\{x_1=x_3=0\}/\C^*$. The weights are $p-1$ and $1$. We expect a $\Z_{p-1}$ quotient singularity, and $\c D_2=\P^1(p-1,1)$. The toric stack method shows this: now $v_2=(0,-1)$, and once again $N_{\c D_2}\iso \Z$. We choose the  isomorphism to be projection on the first coordinate. The $Star$ of $\c D_2$ consists of the cones $\c C_1$ and $\c C_2$. The image of $v_1=(-1,-1)$ in $N_{\c D_1}\iso \Z$ is $-1$. Similarly, the image of $v_4=(p-1,p-1)$ in $N_{\c D_1}\iso \Z$ is $p-1$. Running the BCS construction one indeed obtains a $\P^1(p-1,1)$. 

A similar analysis applies to the $x_4=0$ subspace, and we get another $\P^1(p-1,1)$. In this case the crucial observation is that $\gcd(p-q-2,p-q-1)=1$.

The most interesting case is $\c D_3$, since we start with a toric variety with $N=\Z^2$ and no torsion, but $N_{\c D_3}$ will have torsion. More precisely $N_{\c D_3} \iso \Z\oplus \Z_{p-1}$. This case cannot be treated with the ``naive'' holomorphic quotient method. We start with an explicit isomorphism 
\begin{equation*}
\xymatrix{\Z^2/\langle v_3\rangle \ar[rr]^\phi &&\Z\oplus \Z_{p-1}	}
\end{equation*}
Since we have to quotient by $v_3=(p-1,p-1)=(p-1)\cdot (1,1)$ first we construct a morphism $\psi\!:\Z^2\to\Z^2$ that takes $(1,1)$ to $(1,0)$. We can be quite general here, and for any fixed $a\in \Z$ we can consider 
\begin{equation*}
\psi\!: \xymatrix{\Z^2 \ar[rrr]^{\left(\!\!\!\!
\begin{array}{cc}
a+1&-a\\
-1&1
\end{array}\!\!\!\!\right)} 
&&&\Z^2	}
\end{equation*}
Note that  $\psi$ is an isomorphism, since $\det(\psi)=1$, and is simply a change of basis. It is obvious that $\psi(v_3)=(p-1,0)$, $\psi(v_2)=(a,-1)$ and  $\psi(v_4)=(p-q-a-2,1)$. Therefore the matrix whose kernel gives the $\Z_{p-1}$ action is
\begin{equation*}
\left(\!\!
\begin{array}{ccc}
 a & p-q-a-2& p-1\\
-1& 1	    & 0
\end{array}
\!\!\right)
\end{equation*}
The kernel is generated by the vector $(\alpha,\alpha,\gamma)$, where $(p-q-2)\alpha +(p-1)\gamma=0$. Let $g=\gcd(p-q-2,p-1)$. Then $\alpha =(p-1)/g$ and $\gamma=-(p-q-2)/g$ is a solution. Following \cite{BorisovDM}, $\alpha $ determines the $\Z_{p-1}$ action, and  $\c D_3=\P^1/\Z_{p-1}$. The $\Z_{p-1}$ action is $(x_0,x_1)\mapsto (\xi_{p-1}^\alpha x_0,\xi_{p-1}^\alpha x_1)$, where $\xi_{p-1} $ is a $p-1$st root of unity. We observe that the subgroup $\Z_\alpha$ of $\Z_{p-1}$ generated by $g$ acts trivially. Therefore $\c D_3$ has a generic stabilizer. Note that if $g=\gcd(p-q-2,p-1)=1$, then $\Z_{p-1}$ itself fixes every point. 

To further our understanding of the $X_{p,q}$ geometry, we need to look at the intersection products. Lemma~5.1 of \cite{BorisovDM} shows that the Chow rings of the Deligne-Mumford toric stack and its coarse moduli space (i.e., the singular toric variety) are isomorphic. The intersection products can be computed from the fan (\ref{e:weights}). For reasons to become clear soon, let us introduce the following notation:
\begin{equation*}
f=D_4,\qquad s=D_3.
\end{equation*} 
In these terms $D_2=f$ and $D_1=(p-1)s+(p-q-2)f$ and the intersection products are:
\begin{equation*}
f^2=0,\qquad s^2=-\frac{p-q-2}{(p-1)^2},\qquad s\cdot f=\frac{1}{p-1}.
\end{equation*} 

The two $\C^*$ actions in (\ref{c1}) indicate that $X_{p,q}$ is only a projective bundle over  $\P^1$ rather than a direct product. The fibers are weighted projective lines $\P^1(p-1,1)$, and two of them are  torus invariant: $D_2$ and $D_4$. This structure 
is in line with $f=D_2\sim D_4$ and $f^2=0$. $D_1$  and $D_3$ are sections of this projective bundle. $D_1$ goes through  smooth points in the fibers $\P^1(p-1,1)$, while  $D_3$ goes through the ``singular'' point of every fiber.\footnote{As a scheme $\P^1(p-1,1)$ is just a $\P^1$, but as stacks they are different.} We summarize this structure in the following diagram, where $\pi$ denotes the projection:
\begin{equation*}
\xymatrix{
 X_{p,q}  \ar[d]_{\pi}& \P^1(1,p-1) \ar[l]^{}\\
 \P^1 &   }
\end{equation*}
Therefore $X_{p,q}$ is a stacky generalization of the Hirzebruch surface $\F_n$.
 
Both $\P^1$ and $\P^1(p-1,1)$  have strong exceptional collections. One might hope that $X_{p,q}$
does as well. 

%%%%%%%%%%%%%%%%%%%%%%%%%%%%%%%%%%%%%%%%%%%%%%%%%%%%%%%%%%%%%%%%
\subsection{An exceptional collection on $\F_n$}    \label{s:collfn}

To get a feel for the $X_{p,q}$ exceptional collection it is useful to study the same question on the Hirzebruch surface $\F_n$. $\F_n$ is a $\P^1$-bundle over $\P^1$, with $c_1=-n$. It also equals the projectivization of the rank two split-bundle on $\P^1$: $\P(\O\oplus \O(-n))$. Let $f$ denote the generic fiber, and let $s$ denote the $-n$ section. One has the intersection products: $f^2=0$, $s\cdot f=1$ and $s^2=-n$. 

The Hirzebruch surface $\F_n$ has a strong and complete exceptional collection (see, e.g., \cite{Rudakov:Book}, page 100):
\begin{equation*}
\O\,,\; \O(f)\,,\; \O(s+n f)\,,\; \O(s+(n+1)f)\,,
\end{equation*}
for any $n\in \Z_+$. 

It is instructive to look at the proof of the  strong exceptionality in  \cite{Rudakov:Book}. Our proof in the stacky case will be a more complicated version of this proof, which in the $\F_n$ case is in fact  shorter than the one given in  \cite{Rudakov:Book}, but uses more technology.

%%%%%%%%%%%%%%%%%%%%%%%%%%%%%%%%%%%%%%%%%%%%%%%%%%%%%%%%%%%%%%%%
\subsection{The exceptional collection}    \label{s:coll}

\begin{thm}
For $k_i=\lceil{\frac{(p-q-2)i}{p-1}}\rceil$ and $p-q-2>0$ the following:\footnote{The {\em round up} $\lceil x\rceil$ of a real number $x$ is the smallest integer at least as large as  the number itself.}
\begin{equation} \label{robertscoll}
\begin{split}
\O_{\c X},  \O_{\c X}(D_4) , &\O_{\c X}(D_3+k_1 D_4) , \O_{\c X}(D_3+(k_1+1) D_4) , 
\O_{\c X}(2 D_3+k_2 D_4) , \O_{\c X}(2 D_3+ (k_2+1)D_4) , \\ 
&,\ldots , \O_{\c X}((p-1)D_3+(p-q-2) D_4) , \O_{\c X}((p-1)D_3+(p-q-1) D_4)
\end{split}
\end{equation}
is a strong  exceptional collection on $ X_{p ,q}$.
\end{thm}

The theorem will follow from a more general result concerning strong  exceptional collections on stacky Hirzebruch surfaces. Let $\F_n^p$ be a projective space bundle over the projective line $\P^1$, with fibers isomorphic to the weighted  projective line $\P^1(1,p)$, and  twist specified by $n$, as defined in Chapter~14 of \cite{Stacks}. Alternatively, $\F_n^p$ can be defined as a toric stack with $N=\Z^3$ and cone generated by\footnote{The $ X_{p ,q}$ space in this notation is $\F_{p-q-2}^{p-1}$.}
 \begin{equation}\label{e:weights1}
(-1,-1), \, (0,-1),  \, (p,p),  \, (n,n+1).
\end{equation}
Let $f$ denote the generic fiber. $\F_n^p$ has a section $s$ such that
\begin{equation}
f^2=0,\qquad s^2=-\frac{n}{p^2},\qquad s\cdot f=\frac{1}{p}.
\end{equation} 

\begin{prop}\label{1x4}
The collection
\begin{equation}\label{x4}
\O_{\c X}\,,  \O_{\c X}(f) \,, \ldots \,, \O_{\c X}(i s+k_i f) \,,  \O_{\c X}(i s+(k_i+1) f) \,,\ldots \,, \O_{\c X}(p s+n f)\,, \O_{\c X}(p s+(n+1) f)
\end{equation}
where $k_i=\lceil{\ff{n i} p}\rceil$,  is a strong  exceptional collection on $\F_n^p$ for $p> n$. 
\end{prop}

\begin{proof}
Let us start by observing that $\O_{\c X}(u s+v f)$ is an invertible sheaf for arbitrary integers $u$ and $v$, and therefore 
\begin{equation}\label{e:projfo2}
\Ext^a(\O_{\c X}(u_1 s+v_1 f), \O_{\c X}(u_2 s+v_2 f))=\H^a(\c X,\O_{\c X}((u_2-u_1)s+ (v_2-v_1) f))\,.
\end{equation}
This reduces the problem to computing cohomology groups.

First we  show that the collection is exceptional. Consider two members, $\O_{\c X}(i s+l_i f)$ and $\O_{\c X}(j s+l_j f)$, which appear in this precise order in the above list. The $l_i$'s are necessarily of the form $k_i$ or $k_i+1$ (where $k_i=\lceil{\ff{n i} p}\rceil$). There are two cases to consider
\begin{enumerate}
\item  $i<j$,  
\item $i=j$  and  $(l_i,l_j )=(k_i,k_i+1)$. 
\end{enumerate}

Let us consider the second case first. The $\Ext$ groups in question reduce to $\H^a(\c X,\O_{\c X}(-f))$. Now take the short exact sequence (SES)
\begin{equation}\label{e:flee}
\ses{\O_{\c X}(-f)}{\O_{\c X} }{\O_f}\,,
\end{equation}
and remember that the fiber is $f=\P^1(1,p)$. The associated long exact sequence (LES) in cohomology  shows that 
\begin{equation}\label{e:fleex}
\H^a(X,\O_{\c X}(-f))=0,\qquad\mbox{for all $a\in \Z$.}
\end{equation} 

Returning to the first case, $i<j$, we need to compute $\H^a(\c X,\O_{\c X}((i-j)s+ (l_i-l_j) f)$, for $0\leq i < j \leq p$. Therefore $-p\leq i-j <0$. The vanishing of these cohomology groups follows from the first part of the following lemma:
\begin{lemma}\label{l:43}
For  integers $v\in \Z$  and $a\geq 0$ we have
\begin{equation*}
\H^a(\c X,  \O_{\c X}(u s+ v f))=\left\{ 
\begin{array}{lll}
0 				  & & \mbox{\rm for $-p\leq u <0$}	\\
\H^a(\P^1,  \O(v))   & & \mbox{\rm for $  0\leq u <p$}
\end{array}	\right.
\end{equation*} 
\end{lemma}
\noindent We will prove the lemma shortly. 

This Lemma completes the proof of exceptionality. We are left to show that the collection is strongly exceptional as well. Thus we need that
\begin{equation*}
\Ext^a(\O_{\c X}(i s+l_if),\O_{\c X}(j s+l_j f))=\H^a({\c X}, \O_{\c X}((j-i)s+(l_j -l_i) f))=0,
\end{equation*}
for $a>0$, whenever  $i<j$, or  $i=j$  and $(l_i,l_j )=(k_i,k_i+1)$.

In the second case,  $i=j$, we are reduced to $\H^a({\c X}, \O_{\c X}(f))$. Tensoring (\ref{e:flee}) with $\O_{\c X}(f)$, and using the fact that $f\cdot f=0$, we obtain the SES
\begin{equation*}
\ses{\O_{\c X}}{\O_{\c X}(f) }{\O_f}\,.
\end{equation*}
The associated LES shows that $\H^a({\c X}, \O_{\c X}(f))=0$ for $a>0$, and $\H^0({\c X}, \O_{\c X}(f))=\C^2$. 

The first case, $i<j$, is a bit more involved. Since $0\leq i < j \leq p$, then $0<j- i  \leq p$, and we cannot use the lemma directly. We have to treat the cases $0<j- i  < p$ and $0<j- i  = p$ separately. 

{$\mathbf{0<j- i  < p:}$}
Now the 2nd part of the lemma applies, and we have that 
\begin{equation*}
\H^a({\c X}, \O_{\c X}((j-i)s+(l_j -l_i) f))= \H^a(\P^1,  \O(l_j -l_i)) .
\end{equation*}
Thus $\H^2=0$ automatically, while for $\H^1$ we argue as follows. By assumption $l_i$ equals $k_i$ or $k_i+1$, and similarly for  $l_j$. Furthermore, $k_i=\lceil{\ff{n i} p}\rceil$. Therefore, $k_j\geq k_i$ for $j>i$. Then
\begin{equation*}
l_j -l_i \geq k_j-l_i \geq k_j-k_i-1 \geq -1.
\end{equation*} 
But $\H^1(\P^1,  \O(v))=0$ for $v\geq -1$, and we are done with this case.

{$\mathbf{0<j- i  = p:}$}
In this case we are working with the two very first and very last terms in  (\ref{x4}). So $l_i=0$ or $1$, and $l_j=n$ or $n+1$. One observes that $p s + n f$ is linearly equivalent to $D_1$, as in  (\ref{x5}). As we saw $D_1=\P^1$. As a result we have a SES
\begin{equation*}
\ses{\O_{\c X}(-ps-nf)}{\O_{\c X} }{\O_{D_1}}\,,
\end{equation*} 
Since $(p s + n f)^2=n$ tensoring this with $\O_{\c X}(ps+nf)$ gives the SES
\begin{equation*}\label{x6}
\ses{\O_{\c X}}{\O_{\c X}(ps+nf) }{\O_{\P^1}(n)}.
\end{equation*} 
This gives $\H^0(\c X,\O_{\c X}(ps+nf))=\C\oplus \C^{n+1}=\C^{n+2}$ and $\H^a(\c X,\O_{\c X}(ps+nf))=0$ for $a>0$. 

So far this  takes care of the $\Ext$'s between $\O_{\c X}$ and $\O_{\c X}(ps+nf)$, and resp. $\O_{\c X}(f)$ and $\O_{\c X}(ps+(n+1)f)$. We still need to look at the pairs $\O_{\c X}$ and $\O_{\c X}(ps+(n+1)f)$,  and resp. $\O_{\c X}(f)$  and $\O_{\c X}(ps+nf)$. We deal with the latter one explicitly, while the former one has an identical treatment.

Tensoring (\ref{x6}) with  $\O_{\c X}(-f)$ gives the SES
\begin{equation}\label{x7}
\ses{\O_{\c X}(-f)}{\O_{\c X}(ps+(n-1)f) }{\O_{\P^1}(n-1)}.
\end{equation} 
We have already seen in (\ref{e:fleex}) that $\H^a(X,\O_{\c X}(-f))=0$, and therefore the associated cohomology LES completes the proof of the proposition, provided that $n-1\geq -1$. But this was our initial assumption: $n \geq 0$.
\end{proof}

\begin{proof}[Proof of the lemma]
The proof will use Grauert's theorem, the projection formula, and the Leray spectral sequence. First of all, the fibration structure of $\c X$ guarantees that for two integers $u$ and $v$
\begin{equation*}
\O_{\c X}(u s+ v f)=\O_{\c X}(u s)\otimes_{\c X}  \pi^*\O_{\P^1}(v)\,.
\end{equation*}
Therefore the projection formula gives
\begin{equation}\label{e:projfo1}
R^i\pi_* \O_{\c X}(u s+ v f)=R^i\pi_* (\O_{\c X}(u s)\otimes_{\c X}  \pi^*\O_{\P^1}(v))=R^i\pi_*  \O_{\c X}(u s) \otimes_{\P^1} \O_{\P^1}(v)
\end{equation}
Now let us recall the Leray spectral sequence, which for  a map $f\!:X\to Y$, and a sheaf $\c E$ on $X$ reads:
\begin{equation}\label{e:lss1}
E_2^{i,j}=\H^i(Y,\,R^j f_*\,{\c  E})\Longrightarrow \H^{i+j}(X,\, {\c  E})\,.
\end{equation}
Although the usual Leray spectral sequence was derived for topological spaces, it nevertheless extends to stacks, which have only Grothendieck topologies. The key observation here is to remember that the Leray spectral sequence is a consequence of the composition of the total derived functors: $\Gamma$ -- the global sections functor, and $f_*$ -- the direct image functor. In other words: $\Gamma(Y,f_*\c F)=\Gamma(X,\c F)$ and thus $\R\Gamma(Y,\mbox{-})\comp\R f_*=\R\Gamma(X,\mbox{-})$.\footnote{For more details the reader can consult Sec.~10.8.3 of \cite{weibel}.} All the functors involved are well defined for stacks (see, e.g., \cite{Stacks}), and give rise to the Leray spectral sequence. 

Applying the Leray spectral sequence to $\pi\!: \c X \to \P^1$, we have
\begin{equation}\label{e:lss12}
E_2^{i,j}=\H^i(\P^1,\,R^j\pi_*  \O_{\c X}(u s) \otimes_{\P^1} \O_{\P^1}(v))\Longrightarrow \H^{i+j}(\c X,\,  \O_{\c X}(u s+ v f))\,.
\end{equation}

To compute $R^j\pi_*  \O_{\c X}(u s)$ we can use the stacky version of Grauert's theorem (for the original, see, e.g., Corollary III.12.9 in \cite{Hartshorne:}). First recall that $\c X$ is a fibration over $\P^1$, $\pi\!: \c X \to \P^1$, and the fibers are $f=\P^1(1,p)$. Grauert instructs us to pick a point $y\in \P^1$ in the base, look at the fiber $\c X_y$ over $y$, and for a sheaf $\c F$ compute the cohomology groups of the restriction of $\c F$ to $\c X_y$: $\H^i(\c X_y,\c F|_{\c X_y})$. 

In our case $\c F=\O_{\c X}(u s)$. The restriction of this to a fiber  $f$ is a line bundle. The intersection product $s\cdot f=\ff 1 p$ in $\c X$ shows that the restriction of $\O_{\c X}(s)$ to  $f$ has degree $\ff 1 p$, and therefore it is the bundle that one usually calls $\O_{\P^1(1,p)}(1)$.\footnote{For a nice review of the weighted projective lines the reader can consult Section~9 of \cite{Abramov:Vistoli1}.} By the same token $\c F=\O_{\c X}(u s)$ restricts to $\O_{\P^1(1,p)}(u)$. This is good news, since the cohomology of $\O_{\P^1(1,p)}(u)$ is very simple (note that $K_{\P^1(1,p)}=\O(-1-p)$). In particular:
\begin{equation}\label{x1}
\H^i(\P^1(1,p), \O(u))= \left\{ 
\begin{array}{ll}
0 & \mbox{for $-p\leq u<0$, and $i\geq 0$}	\\
0   & \mbox{for $0\leq u$, and $i> 0$ }	\\
\C   & \mbox{for $0\leq u< p$, and $i= 0$ }.
\end{array}	\right.
\end{equation} 
Using the first two lines of (\ref{x1}) Grauert's theorem implies that 
\begin{equation}\label{x2}
R^i\pi_*  \O_{\c X}(u s)=0 \qquad \mbox{for $-p\leq u<0$ and $i\geq 0$; and for $0\leq u$ and $i> 0$}. 
\end{equation}
The third line of (\ref{x1}) and Grauert's theorem  implies that $\pi_*  \O_{\c X}(u s)$ is locally free of rank $1$ on $\P^1$ for $0\leq u< p$, and using an argument similar to one in the proof of Lemma~V.2.1 of \cite{Hartshorne:} one proves that
\begin{equation}\label{x3}
\pi_*  \O_{\c X}(u s) = \O_{\P^1} \qquad \mbox{for $0\leq u< p$}.
\end{equation} 

In the light of these the Leray spectral sequence (\ref{e:lss12}) degenerates. For $-p\leq u<0$ (\ref{x2}) and Leray imply the first part of the lemma. For $0\leq u$ using both (\ref{x2}) and (\ref{x3}) we get the second statement. 
\end{proof}

%%%%%%%%%%%%%%%%%%%%%%%%%%%%%%%%%%%%%%%%%%%%%%%%%%%%%%%%%%%%%%%%
\section{The toric approach}    \label{s:tor}

From a conceptual point of view it would be desirable to use Kodaira vanishing to prove the exceptionality of our collection. But the usual notion of ampleness does not work for stacks. Nevertheless, we can use the coarse moduli space of the stack to harness the power of ampleness. Let's see how this works.

The starting point is Prop.~3.2 of \cite{BorisovDM} which shows that for $\Sigma$ a rational simplicial fan, the associated toric stack $\c X(\Sigma)$ is a Deligne-Mumford stack. Furthermore, Prop.~3.7 of \cite{BorisovDM} shows that the toric variety $X(\Sigma)$ is the coarse moduli space of $\c X(\Sigma)$. 

At this point we could try to use Theorem~2.1 of \cite{Matsuki:Olsson}, which we reproduce here for the convenience of the reader\footnote{We specialized to the case that is most relevant in our context.}
\begin{thm} \label{t:1}
Let $\c X$ be a smooth proper Deligne-Mumford stack of dimension $d$
with projective coarse moduli space $\pi\colon \c X \to X$. Suppose
that $\pi$ is flat and let $\c L$ be an invertible sheaf on $\c X$ such that some power of $\c L$  descends to an ample invertible sheaf on $X$. Then $\H^i (\c X, K_{\c X}\otimes \c L ) = 0$ for $i >0$.
\end{thm}

Instead of using this theorem directly, we find it more convenient in practice to use Lemma~2.3.4 of \cite{Abramovich:Vistoli}, which for a separated stack $\c X$ with coarse moduli scheme $\pi\colon \c X \to X$ states that $\pi_*$ is exact. As we already mentioned, our toric stacks $X_{p,q}$ fulfill these conditions. Using the Leray spectral sequence, the exactness of $\pi_*$ allows us to reformulate the problem in terms of computing cohomologies on a (singular) toric variety $ X$:
\begin{equation*}
\H^i (\c X, \c L ) = \H^i ( X, \pi_*\c L ).
\end{equation*} 
On $ X$ we can use the power of ampleness, and deduce vanishing on the stack. In fact, this lemma
is used to prove Theorem~\ref{t:1}.

%%%%%%%%%%%%%%%%%%%%%%%%%%%%%%%%%%%%%%%%%%%%%%%%%%%%%%%%%%%%%%%%
\subsection{Toric Kawamata-Viehweg vanishing}

First we establish some useful criteria to determine if the higher cohomology groups of an invertible sheaf on a toric variety vanish. Throughout the section $X$ is a complete toric variety and all divisors are T-invariant Weil divisors. $D_i$ is the divisor associated to the ray $v_i$. Our starting point is a toric Kawamata-Viehweg vanishing  theorem \cite{Mustata}:\footnote{A 
Weil divisor $D$ is {\em $\Q$-Cartier} if there exists an integer $n>0$ such that $n D$ is Cartier. Furthermore, $D$ is ample if $n D$ is ample.}
\begin{thm}\label{cvb1}
If there is an $E = \sum_{j=1}^d d_j D_j$ with $d_j \in {\mathbb Q}$, $0 \leq d_j \leq 1$, such that $D+E$ is an ample ${\mathbb Q}$-Cartier divisor, then $\H^i({\mathcal O}_X(D)) = 0$ for all $i \geq 1$.
\end{thm}

In the toric context we can use the $\Psi$ function  \cite{Fulton:T} to test ampleness. For $D = \sum_i a_i D_i$, the function $\Psi_D: \Delta \to {\mathbb R}$ is linear on each cone $\sigma \subset \Delta$ of the fan and is thus determined by the data $\Psi_D(v_i) = -a_i$.
\begin{prop}[page 70 of \cite{Fulton:T}]\label{Fultonprop}
On a complete toric variety, a Cartier divisor $D$ is ample iff $\Psi_D$ is strictly convex.
\end{prop}

We can reformulate the Kawamata-Viehweg vanishing theorem in the following way:
\begin{cor}
Let $X$ be simplicial and $D$ be as above. If there exists an $E = \sum_i d_i D_i$, with $0 \leq d_i \leq 1$ where $d_i \in {\mathbb Q}$, such that $\Psi_{D+E}$ is strictly convex, then $\H^i(X, \O(D)) = 0$ for all $i>0$.
\end{cor}
\begin{proof}
Since $X$ is simplicial, $D+E$ is ${\mathbb Q}$-Cartier, and $n(D+E)$ is Cartier for some $n>0$. From Prop.~\ref{Fultonprop}, $n(D+E)$ is ample iff $\Psi_{n(D+E)}$ is strictly convex.  But $\Psi_{n(D+E)}$ is strictly convex iff $\Psi_{D+E}$ is strictly convex. 
%Now we can apply Prop.~\ref{Fultonprop} again, and finally use Theorem~\ref{cvb1}.
Now the Corollary reduces to Theorem~\ref{cvb1}.
\end{proof}

%%%%%%%%%%%%%%%%%%%%%%%%%%%%%%%%%%%%%%%%%%%%%%%%%%%%%%%%%%%%%%%%
\subsection{Specializing to a toric surface}

From now on $X$ is a complete simplicial toric surface determined by the fan $\{ v_i \in {\mathbb Z}^2: i = 1, 2, \ldots, n \}$ and $D = \sum_{i} a_i D_i$. We choose the $v_i$'s to be ordered counterclockwise around the origin.

\begin{lemma}\label{surfaceconvex}
With the above notation, $\Psi_D$ is strictly convex iff
\[
\langle v_{i-1}, v_i \rangle a_{i+1} + \langle v_{i+1}, v_{i-1} \rangle a_i +
 \langle v_i, v_{i+1} \rangle a_{i-1} > 0 \ 
\]
for all i.  ($\langle\cdot , \cdot\rangle$ is the usual $2\times 2$ determinant formed from the  components of the two vectors, and $v_{n+1}=v_1$.)
\end{lemma}
\begin{proof}
We show that $\Psi_D$ is strictly convex iff this geometric condition holds.

The question is local, and we need to check convexity for three ordered adjacent vectors, say $v_1$, $v_2$ and $v_3$. There are two cases to consider: $v_1$ and $v_3$ are collinear or not.

\noindent {\em Case 1:} ($v_1$ and $v_3$ are not collinear.)
Since we are in $\RR^2$ we have that:
\begin{equation*}
\alpha\, v_1+\beta \, v_3 =v_2\,\qquad \txt{for some $\alpha,\beta\in \RR$}.
\end{equation*} 
This system of two linear equations has the solution
\begin{equation}\label{bn1}
\alpha=\langle v_2, v_3 \rangle /\langle v_1, v_3 \rangle ,\ \beta=\langle v_1, v_2 \rangle /\langle v_1, v_3\rangle.
\end{equation} 
On the one hand $\Psi_D(v_2)=-a_2$, on the other hand ($\alpha+\beta\neq 0$ since the vectors are not collinear)
\begin{equation*}
\dfrac{1}{\alpha+\beta}\,v_2=\dfrac{\alpha}{\alpha+\beta}\, v_1+\dfrac{\beta}{\alpha+\beta} \, v_3,
\end{equation*} 
and strict convexity implies that 
\begin{equation}
\begin{split}\label{bn2}
\dfrac{1}{\alpha+\beta}(-a_2)=&\Psi_D\left(\dfrac{1}{\alpha+\beta}v_2\right) > \\ &\dfrac{\alpha}{\alpha+\beta}\Psi_D(v_1)+\dfrac{\beta}{\alpha+\beta} \, \Psi_D(v_3)
=\dfrac{\alpha}{\alpha+\beta}(-a_1)+\dfrac{\beta}{\alpha+\beta} (-a_3).
\end{split}
\end{equation} 
Substituting (\ref{bn1}) into (\ref{bn2}) and multiplying through by $R \equiv \langle v_1, v_2 \rangle + \langle v_2, v_3 \rangle$ gives the equation in the statement for $v_1$, $v_2$ and $v_3$. (By definition $v_1$ and $v_2$, resp. $v_2$ and $v_3$, generate strongly convex cones, and therefore $\langle v_1, v_2 \rangle>0$ and $\langle v_2, v_3 \rangle>0$.)
It is immediate that the result for $v_1$, $v_2$, and $v_3$ implies strict convexity locally, and once convexity is checked for adjacent triples, since $\Psi_D$ is piecewise-linear, it is true in general.

\noindent {\em Case 2:} ($v_1$ and $v_3$ are collinear).
In this case $\langle v_1, v_3 \rangle =0$ and 
\begin{equation}\label{zxc1}
\alpha\, v_1+\beta \, v_3 =0\,\qquad \mbox{for some $\alpha,\beta\in \RR$}.
\end{equation}
Since $v_1$ and $v_3$ are in adjacent cones, both $\alpha$ and $\beta$ are positive. By rescaling we choose them such that $\alpha+\beta=1$. Strict convexity in this case is equivalent to the following:
\begin{equation*}
0=\Psi_D (\alpha\, v_1+\beta \, v_3) > \alpha\, \Psi_D(v_1)+\beta\, \Psi_D(v_3)=-\alpha\, a_1-\beta\, a_3.
\end{equation*} 
But (\ref{zxc1}) also implies that $\alpha\, \langle v_1, v_2 \rangle+\beta \, \langle v_3, v_2 \rangle =0$. Combining this with the previous inequality gives  the lemma again.
\end{proof}

\noindent {\em\bf Remark:}
For every $v_i$ consider the three dimensional vector $V_i$, whose first two coordinates are those of $v_i$, and third coordinate is $-a_i$: $V_i=(v_i, -a_i)$. The inequality in Lemma \ref{surfaceconvex} is equivalent to the statement that the $3\times 3$ determinant $\langle 
V_{i-1} , V_i, V_{i+1} \rangle <0$.  This determinant gives the oriented area of the parallelepiped defined by the three $V_i$'s. 

%%%%%%%%%%%%%%%%%%%%%%%%%%%%%%%%%%%%%%%%%%%%%%%%%%%%%%%%%%%%%%%%
\subsection{The $Y^{p,q}$ case}

In this section, using the Kawamata-Viehweg vanishing theorem, we prove that the exceptional collection (\ref{x4}) on $Y^{p,q}$ is simple.  Let $X$ be the toric surface $\F^p_n$ from Sec.~\ref{s:coll}. From Lemma \ref{surfaceconvex}, the three conditions $D=\sum_i a_i D_i$ must satisfy for $\Psi_D$ to be strictly convex reduce to\footnote{The first equation in fact appears twice.}
\be\label{eccr}
a_1 p + a_3 > 0 ,\quad (a_2 + a_4)  p > a_3 n ,\;\quad a_1  n + a_2 + a_4 > 0  .
\ee
To apply Kawamata-Viehweg vanishing for ${\mathcal O}(D)$ we add a suitable $E = \sum_i d_i D_i$ and attempt to show that $\Psi_{D+E}$ is strongly convex. 

First we prove that the higher cohomologies between any two elements in (\ref{x4}) vanish. A general element in this collection is of the form $i D_3 + l_i D_4$, where $l_i=\roof{\frac{ni}{p}}$ or $l_i=\roof{\frac{n i}{p}}+1$, and $i = 0, 1, \ldots, p$. We pick two of these, $i D_3 + l_i D_4$ and $ j D_3 + l_j D_4 $. We are interested in the cohomology of ${\mathcal O}(D) = {\mathcal O}((i-j)D_3 + (l_i-l_j)D_4)$. For $E = \sum_i d_i D_i$ the criteria (\ref{eccr}) for ${\mathcal O}(D+E)$ become
\begin{equation}\label{dione}
d_1 p+i-j+d_3 > 0 ,\; (d_2 + l_i-l_j+d_4)  p > (i-j+d_3) n ,\; d_1  n + d_2 +  l_i-l_j+d_4> 0  .
\end{equation} 
The range of $i-j$ is from $-p$ to $p$ and $l_i - l_j$ from $n+1$ to $-n-1$. For $i-j>-p$ we can set $d_1=d_2=d_4=1$ and $d_3=0$, and the first and third inequalities of (\ref{dione}) are clearly true.  Rewrite the second inequality as
\begin{equation*}
l_i-l_j+2> (i-j) \ff n p.
\end{equation*} 
This inequality holds because $0 \leq \roof{x}-x < 1$, and thus $x-y = x - \roof{y} +(\roof{y}-y) < x - \roof{y} + 1 < \roof{x}-\roof{y}+1$, and finally $l_i-l_j\geq\roof{\frac{n i}{p}}-\roof{\frac{n j}{p}}-1$.

Now we can return to the case $i-j=-p$.\footnote{It 
is interesting that both proofs require a separate treatment of these cases.  }
Here we take $d_i=1$ for all $i$.  The first and third inequalities of (\ref{dione}) are satisfied readily, while the second inequality reads
\begin{equation*}
(l_i-l_j+2)p>(-p+1) n \ .
\end{equation*} 
In this case $l_i - l_j = -n+s$ where $s=-1$, 0, or 1.  Thus the second inequality reduces to 
$(s+2)p > n$, which is true for every $s$ since $p> n$.

To prove that this collection is exceptional, we also need to
establish that $\H^0(\calo(D'-D))$ vanishes whenever $D$ occurs later in the collection than $D'$. In general, the dimension of $\H^0$ is the number of lattice points inside a polygon:  
\be
\dim\H^0(\calo(\sum_i a_i D_i)) = \card  \{ u \in \Z^2\colon u \cdot v_i \geq -a_i \},
\label{PDdef}
\ee
where $v_i$ are the rays of the fan, in our case the ones in (\ref{e:weights1}).

For an invertible sheaf $\calo(a_3 D_3 + a_4 D_4)$, the four inequalities defining the polygon are
\be
x+y \leq 0 ,\quad y \leq 0,\quad x+y \geq -\frac{a_3}{p} ,\quad n x + (n+1) y \geq - a_4 \ .
\ee
Let $D=i D_3 + l_i D_4$ and $D'= j D_3 + l_j D_4 $. Since $D$ occurs later than $D'$ we have that $i\geq j$.
For $a_3=j-i<0$ the polygon is clearly empty.  For $a_3=j-i=0$ necessarily $a_4=-1$.  In this case the first and third inequalities imply that $x+y=0$.  Inserting this result into the second and fourth inequalities yields the contradiction $0 \geq y \geq 1$.
% and this gives a contradiction with the last inequality. 
This completes the proof of strong exceptionality.

The fact that the collection is simple follows from the following proposition:\footnote{The fact that $\F^p_n$ is ample for $n<2p$ follows easily from (\ref{eccr}), since $-K=\sum D_i$, and thus $a_i=1$.}

\begin{prop}
Let $\cale$ be a strong exceptional collection of invertible sheaves on a Fano variety $X$.  If there exists a divisor $E$, satisfying the criteria of the Kawamata-Viehweg vanishing theorem, and $D-D'+E$ is ample for all $D,D'\in\cale$, then $\cale$ is simple. 
\end{prop}
\begin{proof}
For $\cale$ to be simple, we need $\H^q(X, \calo(D-D'-pK)) = 0$  for $q>0$, $p \geq 0$ and all $D,D'\in\cale$.  By assumption $D-D'+E$ is ample, and $-K$ is ample since $X$ is Fano. Therefore $\Psi_{D-D'+E}$ and $\Psi_{-K}$ are both strictly convex, and hence $\Psi_{D-D'+E-pK}=\Psi_{D-D'+E}+p\Psi_{-K}$ is also strictly convex for all $p\geq 0$.  Thus $D-D'+E-pK$ is ample and the higher cohomology of $\calo(D-D'-pK)$ vanishes by Kawamata-Viehweg.  
\end{proof}

%%%%%%%%%%%%%%%%%%%%%%%%%%%%%%%%%%%%%%%%%%%%%%%%%%%%%%%%%%%%%%
\section*{Acknowledgments}

It is a pleasure to thank Paul Aspinwall, Tom Bridgeland, Matthew Ballard, Charles Doran, Mike Douglas, Josh Kantor, Alastair King, Tony Pantev and Paul Smith for useful conversations.  We would also like to thank the KITP of UCSB for providing a stimulating environment where part of this work was done.
The mathematical ``experiments'' that led to the discovery of the exceptional collection (\ref{robertscoll}) were done with the software system Macaulay 2  \cite{M2}. We thank Mike Stillman for sharing his insights with us.
RLK was supported in part by the DOE grant DE-FG02-96ER40949, and by the NSF under Grant No. PHY99-07949.   CPH was supported in part by the U.S. Department of Energy under Grant No.~DE-FG02-96ER40956.

%%%%%%%%%%%%%%%%%%%%%%%%%%%%%%%%%%%%%%%%%%%%%%%%%%%%%%%%%%%%%

\end{document}